\documentclass[11pt,a4paper]{article}
\usepackage[latin1]{inputenc}
\usepackage[english]{babel}
\usepackage{amsmath}
\usepackage{amsfonts}
\usepackage{amssymb,amscd}
\usepackage{makeidx}
\usepackage{graphicx}
\usepackage{enumerate}
\usepackage{tikz-cd}
\usepackage[left=3cm,top=3cm,right=3cm]{geometry}

\newcommand{\bra}[1]{\left[ #1  \right]}
\newcommand\croc[1]{\langle #1 \rangle}
\newcommand\Av[1]{\bra{\croc{ #1 }}}
\newcommand{\Cov}[1]{\mathrm{Cov} \bra{#1}}
\newcommand{\pare}[1]{\left( #1 \right)}

\newcommand{\Y}{Y}
\newcommand{\y}{y}
\newcommand\ci{\perp\!\!\!\perp}
\def\G{\mathcal{G}}
\def\SG{\mathcal{S}}
\newtheorem{ThmCount}{Definition}[section]
\newtheorem{theorem}[ThmCount]{Theorem}

\author{Fatihcan M. Atay\thanks{Max Planck Institute for Mathematics in the Sciences, D-04103 Leipzig, Germany. \texttt{atay@member.ams.org}} 
\and 
Sven Banisch\thanks{Max Planck Institute for Mathematics in the Sciences, D-04103 Leipzig, Germany. \texttt{sven.banisch@universecity.de}}
\and
Philippe Blanchard\thanks{University of Bielefeld,
 Department of Physics
  D-33619 Bielefeld, Germany. 
\texttt{philippe.blanchard@uni-bielefeld.de}}
\and
Bruno Cessac\thanks{Inria Sophia Antipolis M\'editerran\'ee, Neuromathcomp project-team, Sophia Antipolis 06902, France. \texttt{bruno.cessac@inria.fr}} 
\and
Eckehard Olbrich\thanks{Max Planck Institute for Mathematics in the Sciences, D-04103 Leipzig, Germany. \texttt{olbrich@mis.mpg.de}}
}

\title{Perspectives on Multi-Level Dynamics
}

\begin{document}

\maketitle

\begin{abstract}
As Physics did in previous centuries, there is currently a common dream of extracting generic laws of nature in economics, sociology, neuroscience, by focalising the description of phenomena to a minimal set of variables and parameters, linked together by causal equations of evolution whose structure may reveal hidden principles. This requires a huge reduction of dimensionality (number of degrees of freedom) and a change in the level of description. Beyond the mere necessity of developing accurate techniques affording this reduction, there is the question of the correspondence between the initial system and the reduced one. In this paper, we offer a perspective towards a common framework for discussing and understanding multi-level systems exhibiting structures at various spatial and temporal levels. We propose a common foundation and illustrate it with examples from different fields. We also point out the difficulties in constructing such a general setting  and its limitations.
\end{abstract}

\section{Introduction} \label{Fatihcan}

It is generally agreed that complex systems are comprised of a large number of sub-components and their interactions. Moreover, they often exhibit structures at various spatial and temporal levels. 
As a somewhat extreme example, spanning length and time scales of vastly different magnitudes, one can cite the hierarchy of molecules, neurons, brain areas,
brains, individuals, social organizations, economies, etc., which can be viewed as manifestations of the same collective physical reality at different levels. 
Scientific disciplines like biology, neuroscience, psychology,
sociology, economy, and so on, have typically evolved based on notions and descriptions relevant for a certain level. Nevertheless, even within a single discipline it is sometimes desirable to distinguish and investigate several levels and their interactions, such as in the fields of macro and micro economics.
It is therefore a question of both theoretical and practical interest how different descriptions of the same large system at various levels are related to each other. 

\begin{figure}[tb]
\centering
\begin{tikzpicture}
\node(A){$Y$};
\node(B)[node distance=3.5cm,right of=A]{$Y$};
\node(C)[node distance=2.5cm,below of=A]{$X$};
\node(D)[node distance=2.5cm,below of=B]{$X$};
\draw[dashed, ->](A)to node [above] {$\psi$}(B);
\draw[->](C)to node [above]{$\phi$}(D);
\draw[->](C)to node [left]{$\pi$}(A);
\draw[->](D)to node [right]{$\pi$}(B);
\end{tikzpicture}
  \caption{\label{fig:main_mathemacs_diagram} Schematic view of a multi-level dynamical system.}
\end{figure}
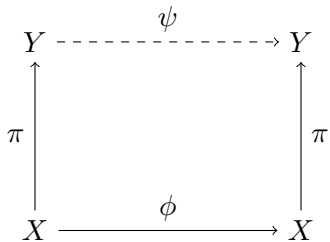

In this paper, we offer some perspectives towards a common framework for discussing and understanding multi-level systems.
Since we cannot hope to address the generality of systems from every area of science, for the presentation we have chosen a few important fields to exemplify the main ideas. 
These are 
information theory (Section~\ref{Eckehard}), Markov chains and agent-based models (Section~\ref{Sven}), mean-field methods in neuroscience (Section~\ref{Bruno}), 
and quantum decoherence (Section~\ref{Philippe}). 
As these examples are very different from each other,
we shall use this introductory section to form a connecting foundation.
The main idea can be graphically illustrated in the diagram of Figure \ref{fig:main_mathemacs_diagram}, which will be at the basis of the discussion in the following sections and will be amended and generalized in various ways.

When we talk about a \emph{system}, we are referring to a \emph{dynamical system}, namely, there is an underlying \emph{state space} $X$ as well as a rule $\phi$ for transition between states. This aspect is represented in the horizontal direction in Figure \ref{fig:main_mathemacs_diagram}. 
The function $\phi$ describes the time evolution in discrete time (such as iteration-based rules like Markov chains) or in continuous time (such as the solution operator or flow of a differential equation), mapping the current state of the system to a future state. The diagram can be replicated in the horizontal direction to correspond to the time trajectory of the system.

The vertical direction of Figure~\ref{fig:main_mathemacs_diagram}, on the other hand, corresponds to the \emph{levels} of the system. Here, one can conceive of another state space $Y$ which describes the system using a different set of variables. The function $\pi$ in the diagram represents the passage from one set of variables to another.
Probably the foremost striking feature in the hierarchy of levels is the (usually huge) difference in the number of variables, i.e., the degrees of freedom, that are used for the description of the system.
Hence, for our purposes, $\pi$ is \emph{not} a coordinate transformation; in fact, it is many-to-one, although it can be assumed to be surjective with no loss of generality.
Correspondingly, one can then refer to the elements of $X$ and $Y$ as micro and macro variables, respectively. 
Such operators as $\pi$ have been studied in the literature  
under the more or less related names of \emph{aggregation}, \emph{reduction}, \emph{projection}, \emph{coarse-graining}, or \emph{lumping}. For instance, $\pi$ may  describe a grouping of several variables into one, which would relate it to the aggregation operation in Markov chains (see Section \ref{Sven}), or it may be a simple averaging operation of the system variables, which would relate it to the \emph{mean-field} methods of reduction (see Section \ref{Bruno}).
The diagram can be repeated in the vertical direction to describe a hierarchy of multiple levels in the system.

The main thread of our discussion hinges upon the question whether the horizontal and vertical directions in Figure~\ref{fig:main_mathemacs_diagram} can be reconciled in some sense.
Formally, this happens if there exists a function $\psi:Y\to Y$ such that the diagram commutes, that is $\pi\phi=\psi\pi$.  
If that is the case, then $\psi$ provides a description of the time evolution of the system using only the $Y$ variables. In other words, the macro variables of $Y$ afford a closed and independent description of the system, at least as far as its dynamical evolution is concerned; and hence we formally view $Y$ as a \emph{level} in the system.
The conditions for $Y$ to constitute a level thus depends on the properties of the dynamics $\phi$ and the reduction operator $\pi$. Such conditions have been derived for several classes of systems;  for more details, the interested reader is referred to the recent preprints \cite{Atay-Roncoroni,Horstmeyer-Atay} and the references therein. 

At this point perhaps some comments on terminology are in order, since there are variations in usage in different fields. 
In this paper, we distinguish between the concepts of \emph{scales} and \emph{levels}. Broadly speaking, we view scales as being essentially
related to the measurement process and the representation of  observed data at different resolutions.

In contrast, we characterize a level by the fact that it admits a closed functional description
in terms of concepts and quantities intrinsic to that level, at least to a certain degree of approximation. 
Thus, when focusing at a particular level, the system function is described and explained in terms of concepts
specific to a certain view on the system.
The distinction of terminology, while not standard in every field, will be important for elucidating the upcoming discussion.

The perspectives about multi-level structures that will be discussed in the following pages can essentially be traced to  
different ways of interpreting the commutativity 
of the diagram in Figure~\ref{fig:main_mathemacs_diagram}. 
In fact, an important case  is when the diagram does \emph{not} commute for given $(\phi,\pi)$, at least not exactly.  
This naturally leads to the concept of \emph{approximately closed} levels and one is then interested in quantifying the discrepancy in terms of \emph{closure measures}. 
For instance, in normed spaces the difference $\Vert \pi\phi-\psi\pi \Vert$, or more precisely,
\begin{equation}   \label{discrepancy}
 \delta = \sup_{x \in \mathcal{A}}  \Vert \pi(\phi(x)) - \psi(\pi(x)) \Vert
\end{equation}
where $\mathcal{A}\subseteq X$ is some subset of states of interest, would be such a measure.
Other measures based on an information-theoretic point of view will be discussed in Section~\ref{Eckehard}.

The lack of exact commutativity can manifest itself in various ways in the description of the system at the macro level $Y$.
In some instances, one has to deal with \emph{memory effects}, 
as discussed in Section~\ref{Sven}, which may require that the 
 state space $Y$ be appropriately extended to obtain a well-defined dynamical system at the macro level. 
In other cases, the loss of information about the exact state of the system, caused by the many-to-one nature of the mapping $\pi$, may typically lead to stochasticity in the description
$\psi$ even if the micro-level dynamics $\phi$ is completely deterministic. However, the converse effect is also possible, as the stochastic effects that may be present at the micro level may be averaged out by the mapping $\pi$, leading to \emph{less} stochasticity at the macro level, even converging to a deterministic description in appropriate limits.
Such a limiting behavior can be visualized by referring to a sequence of diagrams such as Figure~\ref{fig:main_mathemacs_diagram}. 
For instance, let $\phi_n$ describe the dynamics of an $n$-dimensional system on the state space $X_n=\mathbb{R}^n$, 
let $\pi_n:\mathbb{R}^n \to \mathbb{R}$ be the averaging operator, and $Y=\mathbb{R}$. The discrepancy measure given in \eqref{discrepancy}, namely $\delta_n = \Vert \pi_n \phi_n-\psi\pi_n \Vert$, may be nonzero for each $n$ but we may have $\delta_n \to 0$ in the limit as the system size $n\to\infty$.
Section~\ref{Bruno} discusses this aspect in detail in the context of mean field equations. 

Finally, we consider the two examples of multi-level systems where dynamics is specified without reference to a particular equation of motion. 
Section ~\ref{Philippe} considers the problem of quantum decoherence, in which the multi-level description has the quantum and classical representation, and the "quantum level" degrades to the classical one in infinite time.


\section{Information theoretic approach to multilevel systems} \label{Eckehard}

In this section we will provide an information theoretic analysis of multi-level systems as represented in the diagram Fig.~\ref{fig:main_mathemacs_diagram}. We will study, how the information flow within the levels is related to the information flow between the levels and to the existence of closed descriptions within a level.  

In order to so we will consider the following setting:
\begin{itemize}
\item The microscopic state at time $t$ is described by a random variable $X_t$ with discrete states $x \in \mathcal{X}$
\item The macroscopic state at time $t$ is described by a random variable $\Y_t$ with discrete states in $\y \in \mathcal{Y}$.
\item The macrostate $y$ is determined by the microstate via a stochastic map $\pi : X \rightarrow \Y$ representing the aggregation, projection or some other reduction operation. Thus, the choice of $\pi$ defines the macroscopic \emph{level} $\Y$.  
\item The microscopic dynamics $\phi$ is a discrete time dynamics and assumed to be Markov chain with $\phi(x_{t+\tau}|x_t)$  denoting the conditional probability that the system is at time $t+\tau$ in state $x_{t+\tau}$ given that it was in state $x_t$ at time $t$. 
\item As macroscopic dynamics $\psi$ we consider the dynamics ``induced'' by the $\pi$ and $\phi$ in the sense that
\begin{align*}
\psi(\y_{t+\tau}|\y_t):&=p(\y_{t+\tau}|\y_t) \\
&=\frac{\sum_{x_t \in \mathcal{X}} \sum_{x_{t+\tau} \in \mathcal{X}} \pi(\y_{t+\tau}|x_{t+\tau})\phi(x_{t+\tau}|x_t) p(x_t)}{\sum_{x_t \in \mathcal{X}}  \pi(\y_t|x_t) p(x_t)}
\end{align*} 
\item With these definitions the diagram in Fig.~\ref{fig:main_diagram} can be read as a Bayesian network which means that the joint probability distribution factorizes as follows
\begin{equation}
p(\y_{t+\tau},\y_t,x_{t+\tau},x_t)=\pi(y_{t+\tau}|x_{t+\tau})\phi(x_{t+\tau}|x_t)\pi(y_t|x_t)p(x_t) \;.
\label{eq:BayesNet}
\end{equation}
\end{itemize}

Here we made the assumption that the microscopic state provides a complete description in the sense that the dynamics on the microlevel is Markovian, which is a common requirement for defining the ``state'' of system. However, for the sake of simplicity we have additionally restricted ourselves to discrete states and time discrete dynamics. 

For the information theoretic analysis we need the following quantities (for a more detailed treatment see for instance \cite{Cover1991}):

The Shannon entropy \[ H(X)=- \sum_{x \in \mathcal{X}} p(x) \log p(x) \] provides a measure for the uncertainty of the outcome of measuring the random variable $X$. The conditional entropy 
\begin{align*}
H(X|Y) &=- \sum_{x \in \mathcal{X},y \in \mathcal{Y}} p(x,y) \log p(x|y) \\
&=H(X,Y)-H(Y)
\end{align*}
 quantifies the remaining average uncertainty if one knows already the value of $Y$. By already knowing $Y$ the uncertainty of $X$ is reduced by the information that is provided by $Y$, therefore the \emph{mutual information} between $X$ and $Y$ is the difference the entropy $H(X)$ and the conditional entropy $H(X|Y)$:
 \begin{equation}
 I(X;Y)=H(X)-H(X|Y) 
 \end{equation} 
 With the same argument one can also introduce the \emph{conditional mutual information} as the information that $Y$ provides about $X$, if a third variable $Z$ is already known:
 \begin{equation}
  I(X;Y|Z)=H(X|Z)-H(X|Y,Z) 
 \end{equation} 
The microdynamics being Markovian means that the future state $X_{t+\tau}$ is conditionally independent from the past $X_{t-\tau},X_{t-2 \tau},\ldots$ given the present state $X_t$. 
This is equivalent to a vanishing of conditional mutual information 
\[
I(X_{t+\tau};X_{t-\tau},X_{t-2\tau},\ldots|X_t)=0 \;. 
\] 
Thus, the information flow in the lower level is completely characterized by the mutual information of two consecutive time steps $I(X_t;X_{t+\tau})$. In order to see which part of this information can be observed also in the upper level we start form the joint mutual information of the lower and upper level $I(X_{t+\tau},\Y_{t+\tau};X_t,\Y_t)$ and apply the chain rule. The factorization of the joint probability  Eq.~(\ref{eq:BayesNet}) implies the conditional independences \[ X_{t+\tau} \ci \Y_t|X_t \qquad \Y_{t+\tau} \ci (X_t,\Y_t) |X_{t+\tau} \] A simple way to see this is verifying the corresponding d-separation property \cite{Geiger1990} in the Bayesian network. Using that the corresponding conditional mutual informations $I(X_{t+\tau};\Y_t|X_t)=0$  and $I(\Y_{t+\tau};X_t,\Y_t|X_{t+\tau})=0$ vanish one arrives at the following result: 
\begin{align}
I(X_{t+\tau};X_t)&=I(\Y_{t+\tau};\Y_t)+I(\Y_{t+\tau};X_t|\Y_t) \nonumber \\ &+I(X_{t+\tau}:\Y_t|\Y_{t+\tau})+I(X_{t+\tau};X_t|\Y_{t+\tau}, \Y_t)
\end{align}
The single terms on the right side have a clear interpretation: 
\begin{description}
\item[$I(\Y_{t+\tau};\Y_t)$:] Information between two successive steps on the macrolevel. This is part of the information flow in the macrolevel. However, in contrast to the microlevel the macrolevel is not necessarily Markovian and therefore there could be additional contributions to the information flow within this level which we will discuss below. 
\item[$I(\Y_{t+\tau};X_t|\Y_t)$:] Information flow from the micro- to the macrolevel. If this term is non-zero, knowing the micro-state will provide additional information about the future value of the macrostate given that one knows the current value of the macrostate. On the contrary, if this conditional mutual information vanishes, we will say that the macrolevel is {\bf informational closed}. In fact, as we will show below, informational closure implies Markovianity of the macrolevel \cite{Pfante2014a}, but not vice versa, see \cite{Pfante2014} for an example.  
\item[$I(\Y_t;X_{t+\tau}|\Y_{t+\tau})$:] This is also an information flow between the micro- and the macrolevel, but in contrast to the previous one, backwards in time. Here one asks whether knowing the microstate will provide additional information about the previous macrostate which is not known from the current macrostate. While the previous flow is related to information production on the macrolevel, that is which part of the macroscopic randomness can be explained by microscopic determinism, here one deals with the information destruction on the macrolevel and asks whether some of this information survived on the microlevel.  
\item[$I(X_{t+\tau};X_t|\Y_{t+\tau},\Y_t)$] This term contains the part of the microscopic information flow that is irrelevant for the macrodynamics. 
\end{description} 

So far we studied only information flows within a single time step. However, the dynamics on the macrolevel can be Non-Markovian and it is natural to ask, how this Non-Markovianity is related to the information within the macrolevel. Therefore, we extend the basic diagram Fig.~\ref{fig:main_mathemacs_diagram} and consider more time steps. We will use the following notation: $X_{t-n\tau}^t=(X_t,X_{t-\tau},\ldots,X_{t-n\tau})$. In particular, $X_{-\infty}^t$ will denote the complete past.  

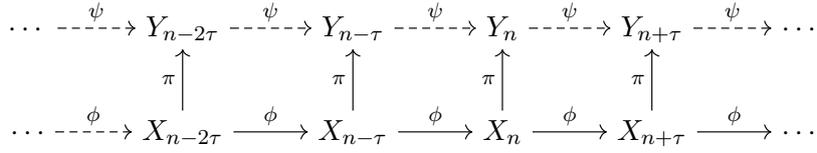
\begin{figure}
\begin{center}
 \begin{tikzcd}
    \cdots \arrow[dashed]{r}{\psi} & \Y_{n-2\tau} \arrow[dashed]{r}{\psi} & \Y_{n-\tau} \arrow[dashed]{r}{\psi} & \Y_n \arrow[dashed]{r}{\psi} & \Y_{n+\tau} \arrow[dashed]{r}{\psi} & \cdots \\
    \cdots \arrow[dashed]{r}{\phi} & X_{n-2\tau} \arrow{r}{\phi} \arrow{u}{\pi} & X_{n-\tau} \arrow{u}{\pi} \arrow{r}{\phi} & X_n \arrow{u}{\pi} \arrow{r}{\phi} & X_{n+\tau} \arrow{u}{\pi} \arrow{r}{\phi} & \cdots
  \end{tikzcd}
\end{center}
  \caption{\label{fig:Markov} Bayesian network of the multilevel system for several consecutive time steps.}
\end{figure}
Then the dynamics on the macro-level is Non-Markovian if and only if  the conditional mutual information 
\[
I(\Y_{t+\tau};\Y_{-\infty}^{t-\tau}|\Y_t) \neq 0 \;.
\]
On the other hand we assumed that the micro-dynamics is Markovian and therefore 
\[
I(X_{t+\tau};X_{-\infty}^{t-\tau}|X_t)=0 \;.
\]
Applying the chain rule to the conditional mutual information  $I(\Y_{t+\tau};X_t,\Y_{-\infty}^{t-\tau}|\Y_t)$ and using the conditional independence  $I(\Y_{t+\tau};\Y_{-\infty}^{t-\tau}|X_t,\Y_t)=0$ yields the following identity:
\[ I(\Y_{t+\tau};X_t|\Y_t)= I(\Y_{t+\tau};\Y_{-\infty}^{t-\tau}|\Y_t)+I(\Y_{t+\tau};X_t| \Y_{-\infty}^t) \]
This identity has the following implication: The forward information flow between the micro- and macrolevel provides an upper bound for the Non-Markovianity of the macrolevel
\begin{equation}
I(\Y_{t+\tau};X_t|\Y_t) \ge I(\Y_{t+\tau};\Y_{-\infty}^{t-\tau}|\Y_t)
\label{eq:InfoInequality}
\end{equation} 
One consequence of this fact is that a vanishing information flow, i.e. \emph{informational closure}, implies Markovianity (i.e. lumpability, see next section) on the macro-level.



\section{Lumpability in Agent-Based Models} 
\label{Sven}

We continue these considerations by applying concepts of the previous sections to a class of models referred to as agent-based models.
In addition to the information-theoretic measures we will focus on the concept of lumpability in Markov chain theory which makes statements about the possibility to aggregate the states of a Markov chain such that the process projected onto that aggregated state space is still a Markov chain.
We start with a description of the microscopic dynamics of a class of agent-based models.

\subsection{A class of agent-based models as Markov chains}

Agent-based models aim at describing certain societal phenomena by putting individual actors (the agents) into a virtual environment, specifying simple rules of how these agents interact and letting the system evolve to observe the macroscopic outcomes of the individual interactions.
A famous example is Schelling's model of spatial segregation \cite{Schelling1971} which shows that a population of agents with a slight preference of settling close to agents of a similar kind produces macroscopic patterns of spatial segregation.
Another often-cited example is the Axelrod model of cultural dissemination \cite{Axelrod1997} emphasizing that a similar mechanism of similarity-driven interaction can provide an explanation for the stable co-existence of populations with different cultural traits.
These models are usually implemented as a computer program and extensive simulation experiments are performed to understand the global outcome of these high-dimensional systems of heterogeneous interacting agents.

Let us consider here a class of models where $N$ agents can be in $\delta$ different states.
Consider further that the agent states are updated sequentially such that from one time step to the other, an agent $i$ is chosen at random along with another agent $j$.
The probability of choosing an agent pair $(i,j)$, denoted here as $\omega_{ij}$, is determined by a weighted interaction network $W$ which defines the neighborhood structure of the agent population.
For instance, the case that two agents are chosen merely at random is encoded by the complete graph $W = K_N$ and the corresponding probability of choosing a pair $(i,j)$ is $\omega_{ij} = 1/N(N-1)$ for all $i \neq j$.
This particular case is sometimes referred to as homogeneous mixing or random mating, depending on the application context.

We consider the case that at one time\footnote{Notice that we use the convention that time indices are in the brackets and the subscript is the agent index. We also say that the process has evolved to time $t+1$ after each interaction event.} step $t$ a single agent pair $(i,j)$ is chosen and only agent $i$ changes its state in dependence of its own current state $x_i(t)$ and the current state of the neighbor $x_j(t)$ by a local update rule
\begin{equation}
x_i(t+1) = u(x_i(t),x_j(t)).
\label{eq:update}
\end{equation} 
This allows us to specify the microscopic transition probabilities between all possible system configurations $x = (x_1,\ldots,x_n) \in \{1,\ldots,\delta\}^N$.
Namely, under sequential update only one agent ($i$) may change at a time giving rise to a transition $x \xrightarrow{i} x'$ such that $x_i \neq x'_i$ and $x_k = x'_k, \forall k \neq i$.
The transition probability for  $x \xrightarrow{i} x'$ is then given by
\begin{equation} 
\phi(x'|x) = \sum\limits_{j : (x'_i,x'_j) = u(x_i,x_j)}^{} \omega_{ij}. 
\label{eq:PhatSSD}
\end{equation}
Under this assumptions, it is easy to show that the micro-level dynamics of such a model can be viewed as a set of random walkers on the Hamming graph $H(N,\delta)$ (with loops).
Hamming graphs are a class of distance-regular graphs in which the nodes are defined as all $N$-tuples of symbols from an alphabet $\{1,\ldots,\delta\}$ and links exist between nodes with a Hamming distance of one.
In our case, the state space of the micro chain is defined by the set of all possible configurations of agents $x = (x_1,\ldots,x_i,\ldots,x_N)$ with $x_i \in \{1,\ldots,\delta\}$ and under sequential update by (\ref{eq:update}) at most one position in the configuration can change from one time step to the next.

As we will see in the next section, the rather regular structure of the micro chain associated to an agent-based simulation model is rather useful for establishing cases in which the macroscopic dynamics induced by $\phi$ and $\pi$ are Markovian.
For this purpose, we next describe the lumpability concept.

\subsection{Lumpability}

Let us restate the (strong) lumpability Theorem 6.3.2 from \cite{Kemeny1976}. 
Let $p(Y | x) = \sum_{x' \in Y}^{} \phi(x'|x)$ denote the conjoint probability for $x \in \mathcal{X}$ to go to the set of elements $x' \in Y$ where $Y \subseteq \mathcal{X}$ is a subset of the configuration space which several micro states are lumped (aggregated) to.
\begin{theorem}
(\cite{Kemeny1976}:124)
A necessary and sufficient condition for a Markov chain to be lumpable with respect to a partition $\mathcal{Y} =  (Y_1,\ldots,Y_r)$ is that for every pair of sets $Y_i$ and $Y_j$, $p(Y_j | x)$ have the same value for every $x$ in $Y_i$.
These common values $\{p_{ij}\}$ form the transition matrix for the lumped chain.
\label{thm:lumpability}
\end{theorem}
In general it may happen that, for a given Markov chain, some projections are Markov and others not.
As different macroscopic properties correspond to different partitions $\mathcal{Y}$ on which the micro process is projected, this also means that it depends on the system property at question whether the time evolution at the associated level of observation is Markovian or not.

On the basis of this Theorem, we have derived in \cite{Banisch2012son} a sufficient condition for lumpability which makes use of the symmetries of the chain: 
\begin{theorem}
\label{thm:symmetry}
Let  $(\mathcal{X},\phi)$ be a Markov chain and $\mathcal{Y} = (Y_1,\dots, Y_{r})$ a partition of $\mathcal{X}$.
For any partition there exists a group ${\G}$ of bijections on $\mathcal{X}$ that preserve the partition $(\forall x \in Y_i$ and $\forall \sigma \in \G $ we have $\hat\sigma(x) \in Y_i)$.
If the Markov transition probability matrix $\phi$ is symmetric with respect to $\G$,
\begin{equation}\label{eq:symmetry_lumpability}
 \phi (x'|x) = \: \phi({\sigma}(x')|{\sigma}(x)) : \forall  {\sigma} \in {\G},
\end{equation}
the partition $(Y_1,\dots, Y_{r})$ is (strongly) lumpable.
\end{theorem}

As an example, let us consider the Land of Oz Markov chain repeatedly considered in \cite{Kemeny1976} (Example 6.4.2).
The idea is that the weather in the Land of Oz is described by the transition probabilities between three different weather states ("Sun", "Rain" and "Snow") as follows
\begin{equation} 
\phi = 
\begin{array}{c}
Sun\\Rain \\Snow 
\end{array}
\left(
\begin{array}{c|cc}
 0 & \frac{1}{2} & \frac{1}{2} \\\hline
 \frac{1}{4} & \frac{1}{2} & \frac{1}{4} \\
 \frac{1}{4} & \frac{1}{4} & \frac{1}{2} \\
\end{array}
\right)  
\stackrel{(strong \ lump)}{\longrightarrow} 
\psi = 
\begin{array}{c}
Nice\\Bad 
\end{array}
\left(
\begin{array}{cc}
0 & 1 \\
 \frac{1}{4} & \frac{3}{4} \\
\end{array}
\right)
\nonumber
\end{equation}
An example for a lumpable partition for this chain is given by the aggregation $\pi$ of the states "Rain" and "Snow" into a macro state "Bad" weather.
The conditions of \ref{thm:lumpability} are satisfied as the transition probabilities from the "Rain" and "Snow" to the lumped state $\{Rain,Snow\}$ as well as to $\{Sun\}$ are equal.
Moreover, this example is also suited to illustrate Theorem \ref{thm:symmetry}.
Namely, it is easy to see that permuting the states "Rain" and "Snow" does not change the transition matrix $\phi$.

\subsection{Application to agent-based models}

Theorem \ref{thm:symmetry} is particularly interesting for agent-based models because it relates the question of lumpability to the \textit{automorphisms} of the micro chain $\phi$ and the structure (or "grammar") $H(N,\delta)$ is known to possess  many automorphisms.
In fact, interpreting $\phi$ as a weighted graph, the symmetry relation (\ref{eq:symmetry_lumpability}) is precisely the usual definition of a graph automorphism.
The set of all permutations $\sigma$ that satisfy (\ref{eq:symmetry_lumpability}) corresponds then to the automorphism group of $(\mathcal{X},\phi)$ and Theorem \ref{thm:symmetry} states that this group can be used to define a lumpable partition.

Now, the automorphism group of the Hamming graph $H(N,\delta)$ is given by the semi-direct product $\SG_{\delta} \rtimes \SG_N$ \cite{Gillespie2013} where the second component accounts for permutation invariance with respect to the agents and the first for symmetries with respect to the $\delta$ possible agent attributes.
For a model that realizes all these symmetries such that $\phi (x'|x) =\phi({\sigma}(x')|{\sigma}(x))$ for all $\sigma \in \SG_{\delta} \rtimes \SG_{N}$, the full automorphism group of $H(N,\delta)$ is realized and this allows for a reduction from a micro chain of size $\delta^N$ to a macro chain with $N/2$ (for even $N$) or $(N+1)/2$ (for odd $N$) states \cite{Banisch2014dnc} -- a quite considerable reduction.

However, the transition probabilities as specified in (\ref{eq:PhatSSD}) satisfy complete permutation invariance with respect to the agents only in the rather particular case of homogeneous mixing where the probabilities $\omega_{ij}$ of choosing agent pair ($i,j$) is equal for all pairs.
Likewise, the permutability of all agent states $\{1,\ldots,\delta\}$ hinges on an update rule $u$ that is unbiased with respect to the different state pairings meaning essentially that exactly the same rule must be used for all pairs of states.
For instance, any dependence or constraint on the distance between $x_i,x_j$ such as assortative mating in population genetics or bounded confidence in opinion dynamics do violate some of these symmetries and, in fact, lead to more interesting macroscopic outcome for this reason.

Therefore, as soon as constraints are implemented in a model (and a model without any constraints is often not that interesting) certain irregularities will appear in the micro chain which reduces the number of automorphisms and require therefore a refinement of the macroscopic partition. 
Interestingly, we can relate constraints due to an heterogeneous interaction graph $W$ and constraints on the rules $u$ independently to the resulting automorphisms of $\phi$.
Namely, let $\SG_{\omega} \subset \SG_N$ denote the automorphisms of $W$ and $\SG_{u} \subset \SG_{\delta}$ the remaining symmetries of the $\delta$ states under $u$, then the automorphism group of $\phi$ is $\G = \SG_u \rtimes \SG_{\omega}$.
The problem of finding a group $\G$ that can be used to construct a lumpable partition by Theorem \ref{thm:symmetry} is therefore reduced to identifying the symmetries of the interaction graph $W$ and the symmetries between the $\delta$ possible states under the update rule $u$ whereas most other approaches to lumpability require the analysis of the $\delta^N$-dimensional micro chain $\phi$.
See \cite{Banisch2015acs} for details and an application to the voter model.

The following conclusions can be drawn: 
\begin{enumerate}
\item 
\emph{The more constrained and heterogeneous the microscopic interaction probabilities and rules, the more irregular the micro process and the lower the chances to obtain a reasonable reduction by lumpability.}
\item
\emph{An observation function $\pi: \mathcal{X} \rightarrow \mathcal{Y}$ will define a lumpable macro process only if it is compatible with the symmetries of the micro chain.}
\item
\emph{
If we decide to stay at the macro level despite the fact that it is not compatible with the symmetries of the micro chain, the macro process is no longer Markovian and non-Markovian memory is introduced at the macro level.}
\end{enumerate}
To illustrate some of these results we will finalize this section with an example.

\subsection{Emergence of memory effects}

The presence of non-trivial temporal correlations is an important empirical fingerprint of data series produced by complex systems.
Lumpability applied to agent-based models makes clear that Markovian macro dynamics can be expected only in exceptional cases.
Far more generally, Markovianity is lost at an aggregate level and a crucial role in this is played by the microscopic heterogeneity implemented in the agent model, the nature of the correlations due to the underlying structure and the constraints in the interaction rules. 
For a better understanding of the contribution of these factors, we envision that lumpability can be combined with the information-theoretic framework for the quantification of "closure" described in Section \ref{Eckehard}. 
In order to better understand the temporal patterns that emerge in the transition from a microscopic model with heterogeneous agents to the macroscopic descriptions of practical interest, the Markovianity measure and information flow (related by (\ref{eq:InfoInequality})) are of particular interest.

\begin{figure}[ht]
\centering
\includegraphics[width=0.75\linewidth]{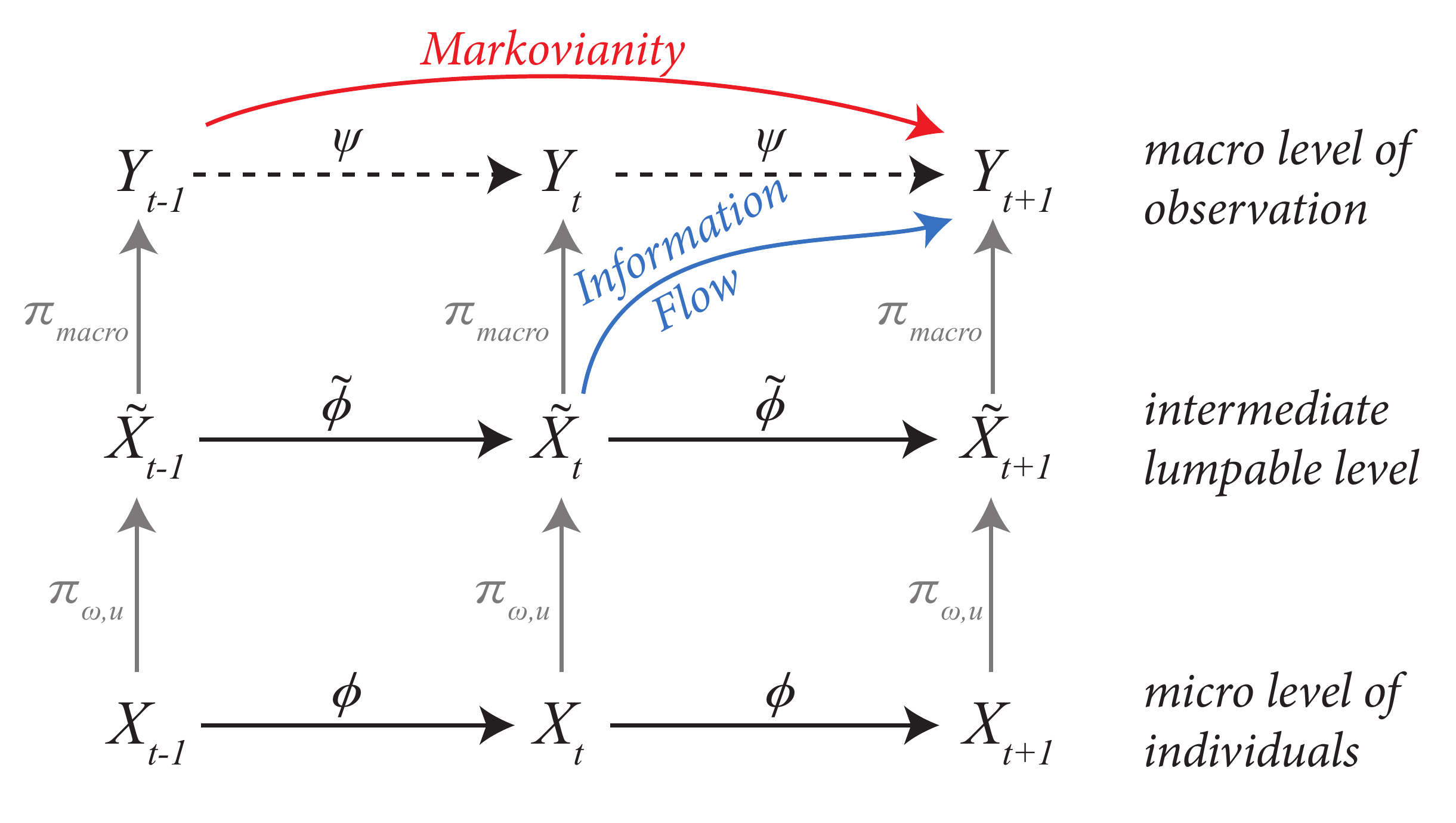}
\caption{Information flow and Markovianity can be computed on the basis of an intermediate meso-level description that complies with $\SG_{u} \rtimes \SG_{\omega}$.}
\label{fig:SingleProjectionMeso}
\end{figure}

For computing information flow and Markovianity for a micro level Markov chain $\mathcal{X}, \phi$ corresponding to an agent-based model and an induced macro-level process $\mathcal{Y}, \psi$ we have to deal with the fact that the size of the micro chain is huge and that the direct computation of these information measures will often be unfeasible.
One way to deal with this is to first use the symmetries of the micro model ($\SG_{u} \rtimes \SG_{\omega}$) and derive a lumpable meso-level description using Theorem \ref{thm:symmetry}.
In Figure \ref{fig:SingleProjectionMeso} this first projection is referred to as $\tilde{\pi}$ and the associated mesoscopic state space is denoted by $\tilde{\mathcal{X}}$.
Since the Markov chain derived for the dynamics at this intermediate level is loss-less with respect to the original micro dynamics, the information quantities involved in the computation of the closure measures can be computed using the reduced-size meso-level chain $(\tilde{\mathcal{X}},\tilde{\phi})$.

This process is particularly applicable to structured populations where a number of homogeneous communities is arranged as a chain or on a lattice.
As an example, we shall consider a voter model on a two-community graph consisting of two homogeneously coupled populations ($a$ and $b$) with strong connections within the sub-populations and weak links connecting the individuals from different populations.
This setup is very similar to the multi-population model of neuronal dynamics dealt with in the next section.
We define the interaction network $W$ as $\bar{W}_{aa} = \bar{W}_{bb} = 1$ meaning that two agents within the same population $a$ or $b$ are linked with weight $1$, and $\bar{W}_{ab} = \bar{W}_{ba} = r$ which, to encode weak ties, is assumed to be smaller than one.

In the model, there are $N$ agents with binary states, $x_i \in \{0,1\}$.
If an agent pair $(i,j)$ is chosen -- and this happens with a probability $\omega_{ij}$ proportional to $W_{ij}$ -- we say that agent $i$ \emph{imitates} the state of $j$ with probability $1-p$ and adopts the \emph{contrarian} view with probability $p$.
Notice that for $p=0$ this corresponds to the voter model.

For binary agent, the microscopic state space is the set of all bit-strings of length $N$ and the graph associated to the micro-level dynamics is the Hamming graph $H(N,2)$, that is: the $N$-dimensional hypercube.
Most typically, the system observable in this type of models is the number of agents in state $1$ given by $k = \sum_{i=1}^N x_i$ also called Hamming weight.
However, due to the inhomogeneity introduced by the fact that not all agent pair are chosen with equal probability (for $\bar{W}_{aa} \neq \bar{W}_{ab}$) the projection onto this partition is not lumpable.

However, in this stylized model, the structure of $W$ is such that it is permutation invariant with respect to permutations of agents within the same sup-population so that a lumpable description is obtained by tracking independently the number of agents in state 1 in the two sub-populations, $k_{a(b)} = \sum_{i \in a (b)} x_i$.
Let $N_{a (b)}$ denoting the number of agents in the two populations, then the state space of this lumped chain is $\tilde{\mathcal{X}} = \{\tilde{x}_{k_a,k_b} : 0 \leq k_a \leq N_a, 0 \leq k_b \leq N_b\}$ which is of size $N_a+1 \times N_b +1$.
It is also clear that this state space is a refinement of the macroscopic partition as $k_a + k_b = k$.

\begin{figure}[ht]
\centering
\includegraphics[width=0.49\linewidth]{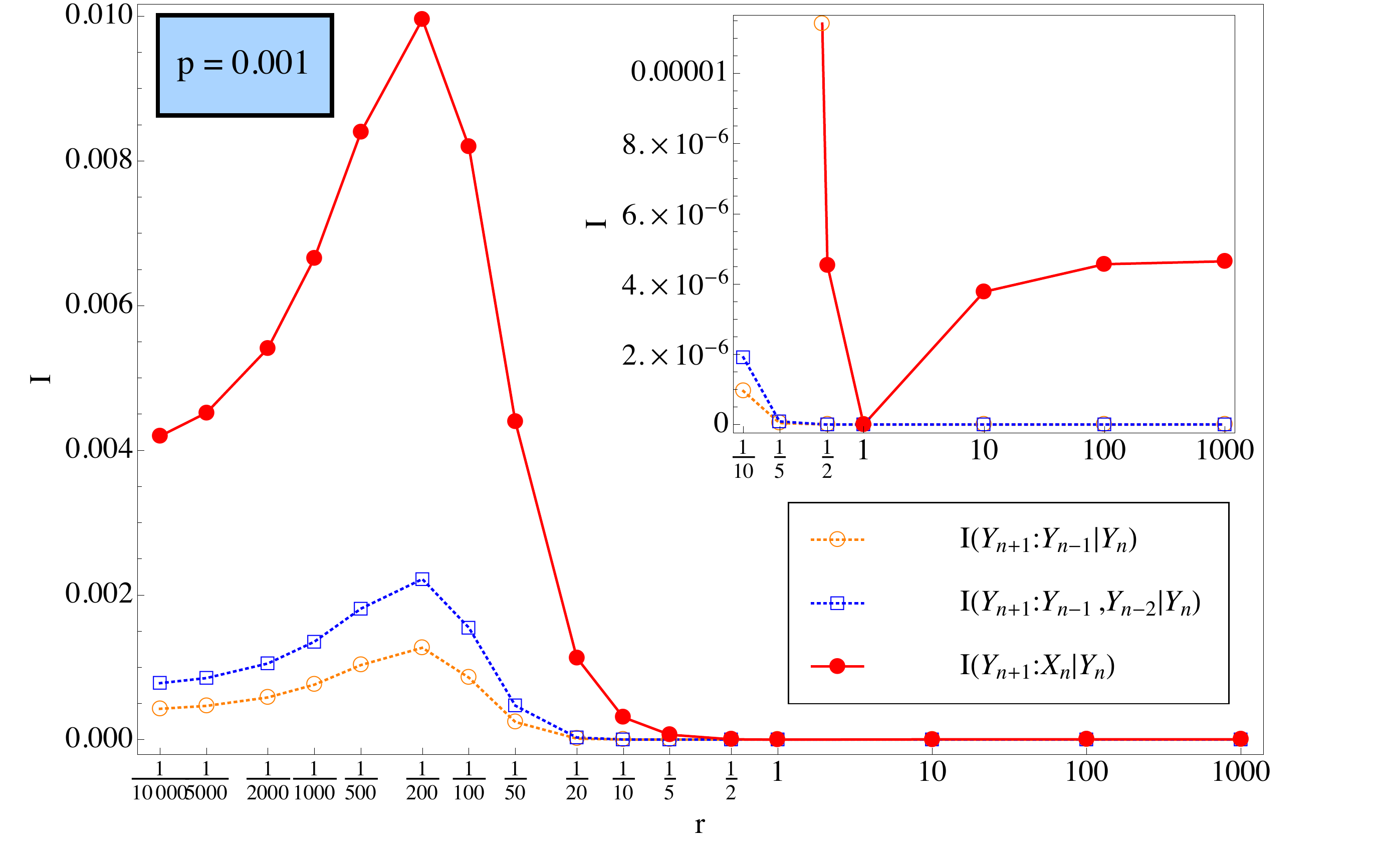}
\includegraphics[width=0.49\linewidth]{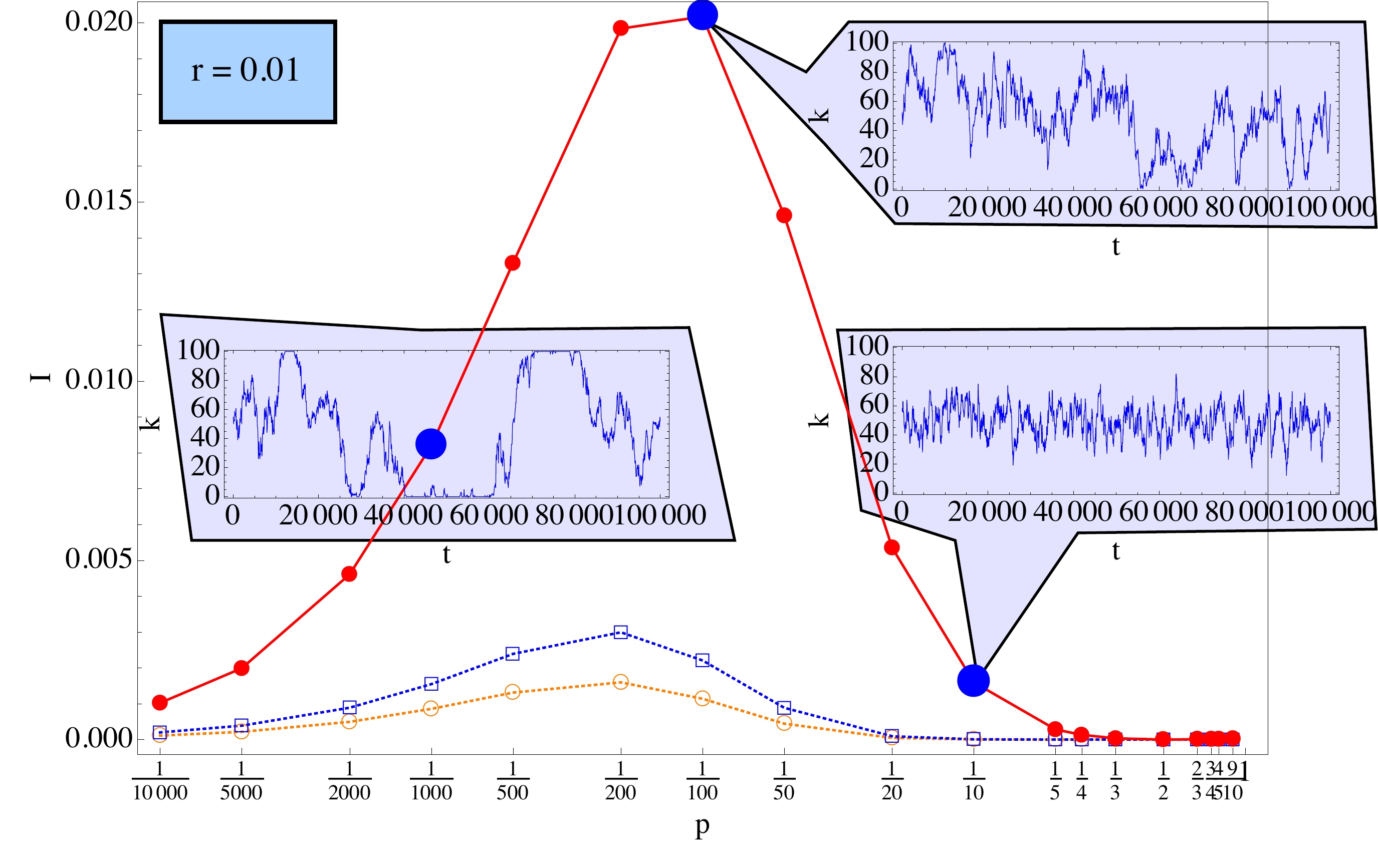}
\caption{Information flow and Markovianity for the two-community voter model. L.h.s.: Closure measures as a function of the ratio $r$ between strong and weak ties. R.h.s.: Closure measures as a function of the contrarian rate $p$ with which agent do not imitate their interlocutor.}
\label{fig:ClosureTwoCom}
\end{figure}

Therefore, on the basis of the meso chain $(\tilde{\mathcal{X}},\tilde{\phi})$ derived for the two-community case, we can exactly compute\footnote{See \cite{Banisch2015springer}, Chapter 7, for all details.} the information flow $I(Y_{t+1};X_t | Y_n) = I(Y_{t+1};\tilde{X}_t | Y_n)$ from the lower to higher level and Markovianity measures for finite histories with $I(Y_{n+1}:Y_{n-1}|Y_n) \leq I(Y_{n+1}:Y_{n-1},Y_{n-2}|Y_n)$.
We show this for a system with $N_a = N_b = 50$ in Figure \ref{fig:ClosureTwoCom}. 
The information flow is shown in red and the Markovianity in orange (one step into the past) and blue (two steps into the past) as a function of the ratio $r$ between weak and strong ties (l.h.s) and the contrarian rate $p$ (r.h.s.).
On the r.h.s. three representative macroscopic realizations of the model are shown to give an idea of the model dynamics.
Notice that the measures vanish in the absence of inhomogeneities ($r = 1$) as shown in the inset on the l.h.s.
Notice also that the information flow from micro to macro is larger than the finite-history Markovianity measure as predicted by (\ref{eq:InfoInequality}).

This demonstrates that the information-theoretic measures described in Section \ref{Eckehard} constitute promising tools to study the relationship between different levels agent-based models. 
Global aggregation over an agent population without sensitivity to micro- or mesoscopic structures leads to memory effects at the macroscopic level.
Future work will have to clarify the range of these memory effects and if the Markovianity approaches information flow in the limit of an infinite history.



\section{Multi levels approach and mean-field methods in neuroscience} \label{Bruno}

In this section we investigate examples of neuronal modelling where the general strategy described in section \ref{Fatihcan} is worth applying. Typically, a neuronal system is composed of several populations of thousands of neurons, connected in a complex way. So, it is tempting to use the mean-field techniques developed in physics, consisting of replacing a population of particles by some quantity (density, field, tensor, order parameter) summarizing
 the relevant properties of this population, to understand the meso- or macro-scopic dynamics.
 That is, one wants precisely to implement the shift in the levels of description described by fig. \ref{fig:main_mathemacs_diagram}. 

As we shall see, it is easy to write down phenomenological equations that perfectly fit in this program and, additionally, meet success when applied to the real brain. However, when one wants to obtain these mesoscopic equations from the microscopic dynamics, some unexpected questions arise, leading to situations where Fig. \ref{fig:main_mathemacs_diagram} or at least its Markovian version Fig. \ref{fig:SingleProjectionMeso} breaks down. Such an examples is fully developed here.

\subsection{The model} \label{Sec:ModelNN}
 In this section we consider a specific class of neural networks model, based on firing rates, introduced by Amari in 1972 and Wilson-Cowan the same year \cite{amari:72,wilson-cowan:72}. The equation of evolution reads:
\begin{equation}\label{DNN}
\frac{d V_i}{dt} = - \frac{V_i}{\tau_i} + \sum_{j=1}^N W_{ij} \, f_j(V_j(t)) + I_i(t) + \sigma \xi_i(t); \quad i=1  \dots N,
\end{equation}
where $V_i(t)$ is the membrane potential of neuron $i$ at
time $t$, $\tau_i$ is the leak rate time constant of neuron $i$. $W_{ij}$ denotes the synaptic interaction weight from $j$ (pre-synaptic neuron) to $i$ (post-synaptic neuron). The $W_{ij}$s do not evolve in time. We have $W_{ij}=0$ when $j$ and $i$ are not connected. $f_j$ is a function characterizing the response curve of neuron $j$, i.e. how does the firing rate of neuron $j$ depends on its membrane potential. We take here a sigmoid, e.g.  $f_j(x)=\frac{1+\text{erf}(\frac{g_j x}{\sqrt{2}})}{2}$. The parameter $g_j$ ("gain") controls the nonlinearity of the sigmoid. $I_i(t)$ represents an external stimulus imposed upon neuron $i$. $\xi_i(t)$ is a white noise, whose amplitude is controlled by $\sigma$. Thus, \eqref{DNN} is a non linear stochastic equation (written in physicists form, for simplicity).\\

Note that eq. \eqref{DNN} is already a mean-field equation as the spike activity has been replaced by the firing rate function $f_j(x)$.\\

A variant of eq. \eqref{DNN} is a multi-population model, where neurons are divided in $P$ populations $a=1 \dots P$, with $N_a$ neurons in population $a$, $N=\sum_{a=1}^P N_a$ \cite{faugeras-touboul-etal:09}. By "population", we mean that neurons can be grouped into subsets having the same characteristics. In the present model, those characteristics are the membrane time constant,
the firing rate function, which depends therefore only on the population (e.g. the gain $g_i=g_a$ on population $a$). We have thus:
\begin{equation}\label{DNNPop}
\frac{d V_i}{dt} = - \frac{V_i}{\tau_a} + \sum_{b=1}^P \sum_{j=1}^{N_b} W_{ij} \, f_b(V_j(t)) + I_a(t) + \sigma \xi_i(t); \quad i \in a, a= 1 \dots P.
\end{equation}

\subsection{The naive mean-field equations} \label{Sec:NaiveMFNN}

We are interested in the mean evolution of membrane potential averaged over neuronal population, in the large number of neurons limit, $V_a(t)= \lim_{N_a  \to  \infty} \frac{1}{N_a} \sum_{i=1}^{N_a} V_i(t)$, assuming this limit exists. To make things simpler in the beginning, assume for a while that $W_{ij}$ only depends on the presynaptic ($b$) and post synaptic ($a$) population, i.e.  $W_{ij} = \frac{\bar{W}_{ab}}{N_b}$, $i \in a, j \in b$, where the scaling factor $\frac{1}{N_b}$ ensures that the sum $\sum_{j=1}^{N_b} W_{ij} \, f_b(V_j(t))$ does not diverge as $N_b \to +\infty$. We obtain:
\begin{equation}\label{dVaEx}
\frac{d V_a}{dt} = - \frac{V_a}{\tau_a} + \sum_{b=1}^P  \bar{W}_{ab}  \, \phi_b(t) +  I_a(t) ;  a= 1 \dots P,
\end{equation}
with:
\begin{equation}
\phi_b(t)= \lim_{N_b \to  +\infty} \frac{1}{N_b} \sum_{j=1}^{N_b} \, f_b(V_j(t)),
\end{equation}
and where we have used $\lim_{N_a  \to  +\infty} \frac{1}{N_a} \sum_{i=1}^{N_a}  \xi_i(t)=0$ (almost-surely).

Eq. \eqref{dVaEx} almost retains the structure of eq. \eqref{DNN}, at the level of populations
(mesoscopic level). However, the big difference is obviously the function $\phi_b(t)$
averaging the non linear influence of population $b$ on population $a$ activity.
In order to have exactly the same structure we would like to write something like
$\phi_b(t)= f_b(V_b(t))$ giving what we call the "naive-mean field equations":
\begin{equation}\label{NaiveMF}
\frac{d V_a}{dt} = - \frac{V_a}{\tau_a} + \sum_{b=1}^P  \bar{W}_{ab}  f_b(V_b(t)) \,  +  I_a(t) ; \quad a= 1 \dots P.
\end{equation}

In this case, the mean-field equation, at the level of average membrane potential population has the same form as the initial equation, at the microscopic level of membrane potentials of neurons. This is a simple example that fits
with fig. \ref{fig:main_mathemacs_diagram} and \ref{fig:SingleProjectionMeso} (in the continuous time case).

\subsection{The commutation assumption} \label{Sec:ComAssumpNN}

Inspected from the mathematical point of view this corresponds to assuming that:
\begin{equation}\label{Commut}
\lim_{N_b \to  +\infty} \frac{1}{N_b} \sum_{j=1}^{N_b} \, f_b(V_j(t)) =  f_b\left( \lim_{N_b \to  +\infty} \frac{1}{N_b} \sum_{j=1}^{N_b} \,V_j(t) \right),
\end{equation}
a commutation property which is wrong in general, as soon as $f_b$ is non linear. Even when $V_j$s are random i.i.d. variables does this assumption fail. A trivial case where it holds however is when all $V_j(t)$ have the same value. 

However, many mean field equations dealing with average activity at the level of populations use the commutation of sigmoidal nonlinearity and membrane potential average. Examples are Jansen-Ritt equations for cortical columns \cite{jansen-rit:95,faugeras-touboul-etal:09} \cite{freeman:75}. Although
these equations are posed ad hoc without mathematical justification, assuming that they hold at the mesoscopic level and, at the same time, that eq. \eqref{DNN} holds true at the neuronal microscopic level somewhat implies
the assumption of commutation. 

As \eqref{Commut} holds at least when all $V_j(t)$ have the same value, trivially equal to the mean value, it is reasonable to guess that \eqref{Commut} is broken down by fluctuations about the mean. How these fluctuations modify the evolution equation, and giving one example where the correct form of mean field dynamics at the mesoscopic level is known, is the topic of this section.  

\subsection{Random synaptic weights} \label{Sec:SynNN}

We are going to obtain the correct mean field equations in the following case. In section \ref{Sec:NaiveMFNN} we have considered that $W_{ij}$ is depending only on the population of pre- and post-synaptic neurons, with a value $\frac{\bar{W}_{ab}}{N_b}$. Here, we extend this situation considering $W_{ij}$ as \textit{independent} Gaussian random variables whose law depends only
on $a$ and $b$, where $W_{ij}$ has mean $\frac{\bar{W}_{ab}}{N_b}$ and variance $\frac{\sigma^2_{ab}}{N_b}$, for $i \in a, j \in b$. Thus, we keep the idea of having an average connectivity strength $\frac{\bar{W}_{ab}}{N_b}$ between population $b$ and $a$, but we now allow fluctuations about the mean and we suppose these fluctuations are uncorrelated. The scaling of the variance ensures
that the term $\sum_{b=1}^P \sum_{j=1}^{N_b} W_{ij} \, f_b(V_j(t))$ in eq. \eqref{DNNPop} has finite and non vanishing fluctuations in the limit $N_b \to +\infty$. A scaling $\frac{1}{N_b^2}$ leads actually to naive mean-field equations since in this case all $V_i$s follow
the same trajectory \cite{amari:72,geman:82}. 

The goal now is to obtain a description of the average behaviour of \eqref{DNNPop}, in the limit $N_b \to +\infty$, $b=1 \dots P$ (thermodynamic limit), where the average is taken both on white noise and synaptic weights distribution (quenched average).  We note
$\bra{ \, }$ the expectation w.r.t. weights and $\croc{ \, }$ the expectation w.r.t. noise.

Taking this average is somewhat the easiest thing to do, in a probabilistic sense (although already quite complex). Indeed, considering the behaviour of averages when taking the infinite size limit corresponds to weak-convergence. Stronger results would deal with almost-sure convergence, namely the typical (on a full measure set) behaviour of a given network in the thermodynamic limit. Such results require however large deviations theory and will not be addressed here. See \cite{moynot-samuelides:02,faugeras-maclaurin:13}.

This case  has been studied, for the first time, by Sompolinsly and co-workers in the model \eqref{DNN}, without noise, without external current, with $f(x)= \tanh(gx)$ which introduces a convenient symmetry $x \to -x$ in the problem ($0$ is always a fixed point). Extensions to discrete time model with broken symmetry $x \to -x$ has been done in \cite{cessac-doyon-etal:94,cessac:94,cessac:95} whereas a multi populations model has been considered in \cite{faugeras-touboul-etal:09}. 

\subsection{Methods to obtain mean-field equations} \label{Sec:MethDMF}

The exact mean-field equations can be derived in 3 ways:

\begin{enumerate}[(i)]

\item The local chaos hypothesis introduced by Amari in 1972 \cite{amari:72} for the continous time dynamics (\ref{DNN}) and used by Cessac and coworkers for a discrete version \cite{cessac-doyon-etal:94,cessac:94,cessac:95};

\item Using the functional generating approach developed for spin-glasses and used, for the first time, by Sompolinsky and coworkers \cite{sompolinsky-crisanti-etal:88}, for the continuous time dynamics (\ref{DNN}), and Molgedey et al for the model (\ref{DNN}) \cite{molgedey-schuchardt-etal:92};

\item Using the large deviations technique introduced by Guionnet and Ben Arous for spin-glasses \cite{ben-arous-guionnet:95} and used by Moynot and Samuelides \cite{moynot-samuelides:02}
for a discrete version of (\ref{DNN}).

\end{enumerate}

These $3$ methods lead to the same mean-field equations.
The advantage of (i) is to be straightforward, although relying on a quite questionable hypothesis
(explained below). The method (ii) belongs to the standard toolbox of physics and is more natural to physicists. To our best knowledge it is the only method up to now\footnote{In principle, large deviations should also allow this study though this has not been done yet.} allowing to study which mean-field solutions are actually observed in the limit $N \to +\infty$. 
In its most recent version it requires a sophisticated formalism (supersymmetry \cite{bouchaud-cugliandolo-etal:96}) and still relies on questionable assumptions: typically the ad hoc cancellation of an auxilary field used in the computation. The advantage of (iii) is to be rigorous and to extend to the correlated case \cite{faugeras-maclaurin:13}. The price to pay is a rather huge mathematical background. 

In order to rapidly present the mean-field equations in the uncorrelated case, we shall focus here on method (i). 

\subsection{Mean-field equations from local chaos hypothesis}
 
 The local chaos hypothesis assumes that, in the thermodynamic limit, neurons become independent from each others and from the $W_{ij}$s. Actually, a weaker property has been proved in \cite{geman:82,moynot-samuelides:02}, the "propagation of chaos": for any subset of $k$ neurons, with $k$ finite, those neurons become independent in the thermodynamic limit.
Denote:
\begin{equation}\label{Ui}
U_{ib}(t)=\sum_{j=1}^{N_b} W_{ij} \, f_b(V_j(t)); \quad i \in a,
\end{equation}
the
sum of "influences" received by neuron $i \in a$, coming from the synaptic population $b$.
From Central Limit Theorem the first consequence of local chaos hypothesis,  is that, in the limit $N \to +\infty$, 
$U_{ib}(t)$ converges in law to a Gaussian process $U_{ab}$ whose law depends only on population index $a$ and $b$. Moreover, local chaos hypothesis allows to compute easily the mean and covariance of this process.

One ends up with the following conclusions.

\begin{enumerate}

\item In the thermodynamic limit all neurons in population $a$ have a membrane potential with the same probability distribution. We denote this potential by $V_a$. Then, eq. \eqref{DNNPop} becomes, in the thermodynamic limit:
\begin{equation}\label{DMFT_Va}
\frac{d V_a}{dt} = - \frac{V_a}{\tau_a} + \sum_{b=1}^P U_{ab}(t) + I_a(t) + \sigma \xi_a(t); \quad i \in a, a= 1 \dots P.
\end{equation}
where $\xi_a  \sim  \xi_i$.

\item $U_{ab}$ is a Gaussian process with mean and covariance given by:
\begin{eqnarray} \label{Law_Uab}
\Av{U_{ab}(t)} &=& \bar{W}_{ab} \, \Av{f_b(V_b(t))};\\
\Cov{U_{ab}(t) U_{cd}(s)}&=& \sigma^2_{ab} \delta_{ac} \delta_{bd}  \,  \Av{f_b(V_b(t)) f_b(V_b(s))}.
\end{eqnarray}

\item $V_a$ is a Gaussian process with mean and covariance given by:
\begin{eqnarray}\label{CovV}
&&\Av{V_a(t)}=\sum_{b=1}^P \bar{W}_{ab} \, \int_{t_0}^t  \Av{f_b(V_b(t))} e^{-\frac{(t-s)}{\tau_a}} ds 
+\int_{t_0}^t I_a(s) e^{-\frac{(t-s)}{\tau_a}} ds\\
&&\Cov{V_a(t),V_{c}(t')}= \\
&& \delta_{ac} \, \sum_{b=1}^P \sigma^2_{ab} \int_{t_0}^t \int_{t_0}^{t'}    \Av{f_b(V_b(t)) f_b(V_b(s))}  e^{-\frac{(t+t'-s-s')}{\tau_{a}}} ds ds'
+  \, \frac{\sigma^2 \tau_a}{2} \bra{1 - e^{-\frac{2 (t-t_0)}{\tau_{a}}}}, \nonumber
\end{eqnarray}
with $t \leq t'$.

\end{enumerate}

These equations constitute a closed set of self-consistent equations, called mean-field equations.

\subsection{From exact mean-field equations to naive ones}

$V_a(t)$ being Gaussian it is easy to obtain the evolution equation of $\Av{V_a(t)}$.
We have:
\begin{equation}\label{Dyn_Mean_Va}
\frac{d \Av{V_a}}{dt} = - \frac{\Av{V_a}}{\tau_a} + \sum_{b=1}^P \bar{W}_{ab} \, \int_{-\infty}^{+\infty} f_b\pare{h \sqrt{\sigma_b(t)}h + \Av{V_b(t)}} Dh + I_a(t),
\end{equation}
with $Dh=\frac{1}{\sqrt{2 \pi}} e^{-\frac{h^2}{2}}$ and where:
\begin{eqnarray}\label{Var_V}
&&\sigma_b(t)= \nonumber\\
&&\sum_{b'=1}^P \sigma^2_{bb'} \int_{t_0}^t \int_{t_0}^{t}   \,  
\frac{1}{2\pi \det C_{b'}(s,s')}
\bra{\int_{-\infty}^{+\infty} \int_{-\infty}^{+\infty} f_{b'}(u) f_{b'}(v)
e^{-\frac{1}{2} X^\dag C_{b'}^{-1}(s,s') X} dX}
  e^{-\frac{(2t-s-s')}{\tau_{b}}} ds ds' \nonumber\\
&&+  \, \frac{\sigma^2 \tau_b}{2} \bra{1 - e^{-\frac{2 (t-t_0)}{\tau_{b}}}}.
\end{eqnarray}
Here, 
\begin{equation} \label{MatCov}
C_{b'}(s,s') = \pare{
\begin{array}{ccc}
\Cov{V_{b'}(s),V_{b'}(s)} & \Cov{V_{b'}(s),V_{b'}(s')}\\
\Cov{V_{b'}(s'),V_{b'}(s)} & \Cov{V_{b'}(s'),V_{b'}(s')}
\end{array}
},
\end{equation}
with:
\begin{eqnarray}\label{CovVbbp}
&&\Cov{V_b(t),V_{b}(t')}=\\
&& \sum_{b'=1}^P \sigma^2_{bb'} \int_{t_0}^t \int_{t_0}^{t'}   \,  
\frac{1}{2\pi \det C_{b'}(s,s')}
\bra{\int_{-\infty}^{+\infty} \int_{-\infty}^{+\infty} f_{b'}(u) f_{b'}(v)
e^{-\frac{1}{2} X^\dag C_{b'}^{-1}(s,s') X} dX}
  e^{-\frac{(t+t'-s-s')}{\tau_{b}}} ds ds' \nonumber\\
&&+  \, \frac{\sigma^2 \tau_b}{2} \bra{1 - e^{-\frac{2 (t-t_0)}{\tau_{b}}}} \nonumber,
\end{eqnarray}
whereas
$X$ is the vector 
$\pare{\begin{array}{ccc} u\\
v \end{array}}$
and $^\dag$ denotes the transpose.

Inspecting equations \eqref{Dyn_Mean_Va} one sees that the evolution of the average is constrained by the time integrations of fluctuations, and those fluctuations are themselves
constrained by a set of self-consistent equations \eqref{MatCov}, \eqref{CovVbbp}, where the covariance $\Cov{V_b(t),V_{b}(t')}$ \textit{integrates the whole history of correlations from the initial time $t_0$}. As a consequence, this evolution is non-Markovian as it integrates the whole history. In this sense, although eq. \eqref{DMFT_Va} looks like
a dynamical system equations, it is not: the term $\sigma_b(t)$ introduces a whole history dependence of the trajectories.\\

Clearly, although eq. \eqref{Dyn_Mean_Va} looks very similar to \eqref{NaiveMF} they are deeply different. Note, however that if fluctuations vanish ($\sigma_b(t)=0$) both equations are
identical.

\subsection{Mean-Field solutions}

Solving those equations is a formidable task. To our best knowledge it has been achieved in only one case, for one population, $f(x)=\tanh(gx)$, $I_a=0$, no noise, by Sompolinsky et al, \cite{sompolinsky-crisanti-etal:88}. Assuming stationarity, they where able to show the quantity $\Delta(\tau)=\Av{V(t) V(t')}, \tau=t'-t$ obeys the Newton equation:
$$
\frac{d^2 \Delta}{d \tau^2} = - \frac{\partial {\cal V}}{\partial \Delta}
$$
for some potential ${\cal V}$ that can be explicitly written. From this it is easy to draw  the phase portrait and infer the time evolution of $\Delta$. From this analysis these authors were able to show the existence of several dynamical regimes, from stable fixed point, to periodic solutions, to chaos. Moreover, the dynamic mean field method they used,
based on generating functional, allowed to show that periodic solutions are in fact not observable in the thermodynamic limit. Analysis of the finite dimensional system confirms this result: there is transition to chaos by quasi-periodicity when increasing the parameter $g$  where the intermediate phase (periodic orbits and $T2$ torus) occurs on a $g$ range that vanishes as $N \to +\infty$ \cite{cessac-doyon-etal:94,cessac:95}.

\subsection{Discussion}

In this section we have given an example where mean-field equations at the mesoscopic level
can be obtained from microscopic dynamics, by performing a suitable averaging. However,
the structure of mean-field equations is in general quite different from the microscopic ones. Going to the mean-field, one replace $N$ equations by eq. \eqref{DMFT_Va}, \eqref{MatCov}, \eqref{CovV}, but the price to pay is the inheritence of a non Markovian structure, extremely hard to integrate even numerically. Here, averaging over the synaptic weights fluctuations produces therefore a mean-field dynamics rather difficult to interpret, requiring new tools and concepts, in the spirit of those developed for spin-glasses in the eighties, although with a different, non relaxational dynamics \cite{sompolinsky-zippelius:82}. Additionally, introducing weights correlations, which are expected as soon as one e.g. introduces synaptic plasticity complexifies even more the picture \cite{dauce-etal:98,naude-cessac-etal:13,faugeras-maclaurin:13}.

In the context of this paper and of the general scheme of fig. \ref{fig:main_mathemacs_diagram} and \ref{fig:SingleProjectionMeso}, one clearly sees that fluctuations induced by synaptic weights inhomogeneity breaks down the Markov property of the initial equations. At first glance, it looks to be generated by the somewhat artificial procedure of averaging over synaptic weights and noise. This is partly true. However, the problem is deeper. Indeed, as we said above, the averaging corresponds, in a probabilistic context, to a weak form of convergence as the number of neurons tends to infinity. A stronger type of convergence (e.g. almost-sure, i.e. for a measure one set of synaptic weights and noise trajectory selection) would be preferable. Now, if almost sure convergence holds, the solution has to converge to the weak solution, the one we found here. 
In this case, we obtain, that for almost-every realisation of networks and noise, the mean-field solution actually also breaks down the Markovianity.  

As we remarked in subsection \ref{Sec:ComAssumpNN}, the naive mean-field equations are exact when
when all $V_j(t)$ have the same value, a rather exceptional situation. This can however be considered as a good approximation to the case when fluctuations (controlled by the term $\sigma_b$ in \eqref{Dyn_Mean_Va} ) are small. In any other cases, the relevant equations are the non Markovian dynamic mean field equations, which produce quite a non trivial dynamics.

As a final remark, note however that the equations for covariance includes an exponential decay with time, so that a Markovian approximation with time cut-off can easily be proposed.

\section{Quantum decoherence as a multi-level system } \label{Philippe}

The notion of environmental decoherence has been widely discussed and accepted as the mechanism by which classicality emerges in a quantum world. Decoherence explains why we tend not to observe quantum behavior in everyday macroscopic objects. For example, one of the most revolutionary elements introduced into
physical theory by quantum mechanics is the superposition
principle.
If $|1\rangle$ and $|2\rangle$ are two states, then
quantum mechanics tells us that any linear combination
$\alpha|1\rangle + \beta |2\rangle$ 
also corresponds to a possible state. 
Whereas
such superposition of states have been experimentally
extensively verified for microscopic systems,
it is apparently not the case  
of the everyday world -- 
a Schr\"{o}dinger cat that is
a superposition of being alive and dead 
does not bear much resemblance
to reality as we perceive it.
Why does the world appear
classical to us, in spite of its supposed underlying quantum
nature?
Quantum decoherence also explains why we do see classical fields emerge from the properties of the interaction between matter and radiation for large amounts of matter. 

Quantum decoherence can be viewed as the loss of information from a system into the environment (often modeled as a heat bath), since every system is loosely coupled with the energetic state of its surroundings. Viewed in isolation, the system's dynamics are non-unitary (although the combined system plus environment evolves in a unitary fashion). Thus the dynamics of the system alone are irreversible. As with any coupling, entanglements are generated between the system and environment. These have the effect of sharing quantum information with, or transferring it to, the surroundings. Quantum decoherence
represents an extremely fast process for macroscopic objects, since these are interacting with many microscopic objects, with an enormous number of degrees of freedom, in their natural environment.

As we show in this section, it is remarkable that quantum decoherence provides an 
example of a multi-level system, in which 
the time evolution of observables is 
reduced to 
a completely positive dynamical map
under conditional expectation
and then, 
can be replaced by the effective 
 dynamics, 
as time tends to infinity. 
 In this section, 
we follow our joint presentation with Mario Hellmich on decoherence in infinite quantum systems \cite{Blanchard:2010}. 

\subsection{Setting the stage}

In the standard interpretation of quantum
mechanics, a measurable operator in a Hilbert space -- an observable  corresponding to a physical
quantity  -- 
has a definite value if and only if the system
is in an eigenstate of the observable.
If the system
is in a superposition of such eigenstates, 
according to the orthodox interpretation,
it is meaningless to speak of the state of the system as having
any definite value of the observable at all.
In a typical laboratory experiment involving some physical system, 
we can identify two subsequent phases: a preparation 
which is followed by a measurement. 

Following \cite{Schlosshauer:2005}, 
we say that 
a microscopic system $\mathcal{S}$, 
represented by basis
vectors $\left\{|s_n\rangle\right\}$ 
in a Hilbert space $\mathcal{H}_{\mathcal{S}}$, 
interacts 
in the ideal
measurement scheme
with
a measurement apparatus $A$, 
described by basis vectors
$\left\{|a_n\rangle\right\}$ 
 spanning a Hilbert space $\mathcal{H}_{A}$, 
where the $|a_n\rangle$ are
assumed to correspond to macroscopically
 distinguishable
 positions that correspond to the outcome
of a measurement if $\mathcal{S}$ 
is in the state $|s_n\rangle$.
The dynamics of the quantum state of a quantum system
is given by the Schr\"{o}dinger equation.
If $\mathcal{S}$ is in a microscopical
superposition of states
$\sum_{n}c_n|s_n\rangle,$
 and $A$ is in the initial "prepared"
quantum state $|a_r\rangle$, 
the linearity of the Schr\"{o}dinger equation 
entails that the total system $\mathcal{S}A$,
 assumed to be represented
by the Hilbert product 
space $\mathcal{H}_{\mathcal{S}}\otimes \mathcal{H}_A$, 
evolves with time according
to 
$$
\left.\left( \sum_{n}c_n|s_n\rangle\right)\right|a_r\rangle
\rightarrow_t\sum_{n}c_n|s_n\rangle|a_n\rangle
$$ 
where the coefficients $c_n$ are some functions of time. 
This dynamical evolution is often referred 
to as a {\it preparation}
procedure  (or a {\it premeasurement} as in  \cite{Schlosshauer:2005}) 
in order to emphasize that the process
does not suffice to directly conclude
that a measurement has actually been completed.
A preparation  procedure will be denoted by $\varphi$
and a measurement effected by using some instrument will be 
denoted by $A.$ The probability 
that the measurement gives rise to  a value lying in the 
Borel set $E \subseteq \mathbb{R}$
will be denoted by $P[\varphi,A;E].$
The set of measurement is assumed to be discrete, indeed.
Two different preparation procedures $\varphi_1$ and $\varphi_2$
such that the corresponding probability distributions
$P[\varphi_1,A;\cdot]$ and $P[\varphi_2,A;\cdot]$
are identical for any instrument $A$ are said to be equivalent \cite{Araki:1999},
$\varphi_1\sim \varphi_2.$ 
An equivalence class of procedures with respect to the defined 
equivalence relation is called a {\it state}, and the set of all 
states will be denoted by $\Sigma.$
Similarly, if for two instruments $A_1$ and $A_2$ the probability distributions 
$P[\varphi_1,A;\cdot]$ and $P[\varphi_2,A;\cdot]$ 
agree for all states $\varphi\in \Sigma$ we call
the instruments equivalent, 
$A_1\sim A_2,$
and the equivalence classes of this equivalence relation 
are called {\it observables}. The set of all observables
will be denoted by $\mathcal{D}.$
If any measurement of  $A\in\mathcal{D}$  gives only positive results,
$\varphi(A)\geq 0,$ for any $\varphi \in \Sigma$, 
we call $A$ positive, $A\geq 0$.

We further assume that $\mathcal{D}$ can be embedded
in a $C^*$-algebra $\mathcal{A},$
a complex algebra of continuous linear 
operators on a complex Hilbert space,
which is a topologically closed set 
in the norm topology of operators and 
is closed under the operation of taking 
adjoints of operators.
Then the observables correspond
to the self-adjoint $A=A^*$ elements of $\mathcal{A}$.
The states $\Sigma$ are identified with the set of all continuous
positive and normalized functionals on $\mathcal{A}$, 
$\Sigma\cong \left\{ \varphi\in\mathcal{A}^*:\varphi(A^*A)\geq 0, \forall A \in \mathcal{A}, \varphi(\mathbb{I})=1 \right\}$.

The (presumably reversible) time evolution of a closed quantum system described in a certain representation by a von Neumann algebra $\mathcal{M}$ is given by a one-parameter group of ${}^*$-automorphisms $\{\alpha_t\}_{t\in{\mathbb{R}}}$
of $ \mathcal{M}$. That is, each $\alpha_t$ is a bijective linear map on $\mathcal{M}$ such that $\alpha_t(xy)=\alpha_t(x)\alpha_t(y)$ and $\alpha_t(x^*)=\alpha_t(x)^*$ for all $x,y\in\mathcal{M}$, and such that it satisfies the
group property $\alpha_s\circ \alpha_t=\alpha_{s+t}$ for all $s,t\in\mathbb{R}.$ Moreover, we shall assume 
that $t\mapsto \varphi\left(\alpha_t(x)\right)$
is continuous for any normal state $\varphi,$ 
and that expectation values move continuously in time --
the so called {\it weak}${}^*$ {\it continuity}.

\subsection{Open systems and decoherence}

We consider a subsystem of a closed physical system 
described by a von Neumann algebra $\mathcal{N}$
containing the observables of the system, together with a reversible time evolution given by a weak${}^*$ continuous one-parameter group  $\{\alpha_t\}_{t\in{\mathbb{R}}}$ of ${}^*$-automorphisms.
The subsystem will be described by a subalgebra $\mathcal{M}\subseteq \mathcal{N}$ wich includes those observables pertaining to the subsystem. 
In addition we assume the existence of a normal conditional expectation 
$E:\mathcal{N}\to \mathcal{M},$ which 
is a weak${}^*$ continuous linear and idempotent map of norm one.
Then the reduced time evolution is defined by 
$T_t(x)=E\circ\alpha_t(x)$, $x\in\mathcal{M}$, $t\geq 0$. In general, 
the reduced time evolution is no longer reversible,
reflected by the fact that the evolution operators $T_t$ are noninvertible. 
The reduced time evolution 
$\{T_t\}_{t\geq 0}$ is a completely positive linear map for
every $t\geq 0$, with $\|T_t\|\leq 1$, and $t\mapsto T_t(x) $ is ultraweakly 
continuous for all $x\in\mathcal{M}.$

The reduced dynamics  $\{T_t\}_{t\geq 0}$ is said to 
display {\it decoherence} if there is a decomposition 
$\mathcal{M}=\mathcal{M}_1\oplus \mathcal{M}_2$ such that 
for every observable $x\in \mathcal{M}$ there exist a unique decomposition 
into self-adjoint operators $x_1\in\mathcal{M}_1$ and 
$x_2\in\mathcal{M}_2$ such that $x=x_1+x_2$, and 
$\lim_{t\to\infty}\varphi\left(T_t(x_2)\right)=0,$
for all normal states $\varphi,$ i.e. 
all expectation values of $x_2$ converge to 0 as time tends to infinity, 
so that $\mathcal{M}_2$ part is beyond experimental resolution after 
decoherence has taken place. 
Thus, in the limit $t\to\infty$ the system behaves effectively
(and therefore valid for all practical purposes)
 like a closed system described by the von Neumann algebra of effective observables
 $\mathcal{M}_1$
with reversible time evolution given by the one-parameter group 
$\{\beta_t\}_{t\in\mathbb{R}}.$
We summarize the algebraic framework in the 
following diagram.  
\begin{equation}
\begin{CD}
\mathcal{N} @>\{\alpha_t\}_{t\in \mathbb{R}}>> \mathcal{N} \\
@V E VV @VV E V\\
\mathcal{M}_1\oplus \mathcal{M}_2 @>\{T_t\}_{t\geq 0}>> \mathcal{M}_1\oplus \mathcal{M}_2  \\
@V t\to\infty VV @VV t\to\infty V\\
\mathcal{M} _1 @>\{\beta_t\}_{t\geq 0}>> \mathcal{M}_1 
\end{CD}
\label{diag_demand}
\end{equation}
The time evolution of observables 
contained in the von Neumann algebra $\mathcal{N}$
is described by the weak${}^*$ continuous one-parameter group 
of ${}^*$automorphisms $\{\alpha_t\}_{t\in\mathbb{R}}.$ 
Then, under the action of the conditional expectation $E,$
the dynamics is reduced to a completely positive 
linear map $\{T_t\}_{t\geq 0}$.
And, for all practical purposes,
in the limit $t\to\infty,$ 
the decoherent system can be considered 
as a closed system described by the von Neumann
 algebra of effective observables
 $\mathcal{M}_1$
with reversible time evolution
 given by the one-parameter group 
$\{\beta_t\}_{t\in\mathbb{R}}.$
 
It is remarkable that the quantum decoherence diagram shown above constitutes nothing else but a quantum mechanical version of the diagram shown in Figure 1, representing multi-level systems schematically. In fact, our diagram has two levels -- quantum and classical -- instead of the single classical level of the previous sections. Namely, we have shown that a subsystem of a closed physical system described by a von Neumann algebra $\mathcal{N}$ containing the observables of the system,
with a reversible time evolution given by a weak* continuous one-parameter group of automorphisms, can be described by a reduced dynamics -- 
represented by the new set of "macro-variables" analogous to $Y$ in the diagram in Figure 1 --
-- which is said to display decoherence if there is a decomposition of observables into a direct sum of self-adjoint operators belonging to the classical and quantum subalgebras, and all expectation values of the quantum part converge to 0 as time tends to infinity. 

It is also remarkable that our diagram resolves the conundrum question related to the Kochen-Specker (KS) theorem. KS proves that there is a contradiction between two basic assumptions of the hidden variable theories intended to reproduce the results of quantum mechanics: that all hidden variables corresponding to quantum mechanical observables have definite values at any given time, and that the values of those variables are intrinsic and independent of the device used to measure them. This contradiction is caused by the fact that quantum mechanical observables need not be commutative, so that  it turns out to be impossible to simultaneously embed all the commuting subalgebras of the algebra of these observables in one commutative algebra, assumed to represent the classical structure of the hidden variables theory, if the Hilbert space dimension is at least three. We overcome and explain the paradox by showing that all expectations of the non-commutative components corresponding to quantum mechanical observables vanishes with time -- and, in infinite time, only commutative, classical observables can be measured in the macroscopic world.

Furthermore, by looking at the structure of 
the von Neumann
 algebra of effective observables
$\mathcal{M}_1$ and the evolution 
$\{\beta_t\}_{t\in \mathbb{R}},$
we may classify different scenarios of decoherence.

\subsection{Scenarios of decoherence}

In the following list we briefly introduce possible scenarios that can emerge due to decoherence skipping the details for 
\cite{Blanchard:2003}.

 If the von Neumann
 algebra of effective observables $\mathcal{M}_1$ 
is commutative and 
$\{\beta_t\}_{t\in \mathbb{R}}=\mathtt{id}$
for all $t$, then we speak of environmentally 
induced {\it pointer states}. This 
situation is characteristic for a measuring apparatus, 
where the von Neumann
 algebra of effective observables
$\mathcal{M}_1$ contains the observables representing the pointer positions of the apparatus. The commutativity ensures that we obtain a classical probability distribution over the pointer positions whereas the triviality of 
$\{\beta_t\}_{t\in \mathbb{R}}$ ensures that the pointer observables
are immune to the interaction with the environment.

 If the von Neumann
 algebra of effective observables $\mathcal{M}_1$ 
is noncommutative but 
has a nontrivial center --
 the set of all those elements 
that commute with all other elements
-- we speak of environmentally induced
 {\it superselection rules}. 
Then the center of algebra 
contains the superselection observables 
which are classical observables, 
taking a definite value in each sector.
 
 If the von Neumann
 algebra of effective observables $\mathcal{M}_1$ 
is a factor algebra again, then after decoherence 
the system effectively behaves like a 
closed system with evolution
 $\{\beta_t\}_{t\in\mathbb{R}},$ 
still having a pure {\it quantum character}.
However, it may be smaller than the original system.

If the von Neumann
 algebra of effective observables $\mathcal{M}_1$ 
is commutative, we speak of an environment 
induced {\it classical structure}.
 Then the system can effectively be described 
in terms of classical probability. However, a
classical physical system has more structure. 
For example, 
the underlying classical probability 
space and the time evolution, 
given by $\{\beta_t\}_{t\in\mathbb{R}}$, 
need not come from a classical dynamical system, 
or more precisely, 
from the Hilbert space representation 
of a topological or
smooth classical dynamical system 
with a evolution given by a flow 
on phase space $\Omega.$

Finally, if the von Neumann
 algebra of effective observables 
$\mathcal{M}_1$ is a constant (a number on its own), 
the system is ergodic.


\section{Conclusion}
The study of multi-level systems is a challenging endeavor in many ways, from data collection and modeling to analysis and control. 
The individual sections of the present article form a sample of quite different examples coming from various application areas. 
Through these examples we have, on the one hand, aimed to point out the difficulties in constructing a general theory of multi-level systems, while on the other hand we have maintained that a common foundation may be possible upon which such a theory can be built. 

It is hoped that the various perspectives presented here will be useful for a common framework in the discussion of multi-level structures within and across different scientific disciplines.

As discussed in the introduction, the main interest of developing multi-level description of systems is, on the one hand, to simplify the description in the sense of reducing the number of degrees of freedom, and on the other hand, and more fundamentally, to extract from this reduction some emerging principles that were not visible at the initial level. A similar process has occurred several times in the history of physics with e.g. the development of electromagnetism or statistical physics. Certainly, the dream of extracting generic laws of nature in economics, sociology, neuroscience, in the same way as physics did, is part of the motivation for developing multi-level descriptions.

However, there exists strong and structural differences with physical systems. First, when dealing with multi-agents, neuronal systems, economic actors, etc., the nature of interactions is quite different from physics: they are not symmetric, they depend on a possibly very long history (memory), and they can even display a form of anticipation of the future as well as different degrees of expectation (e.g. expectation of expectation \cite{barber-buchinger-etal:06}). Additionally, evolution is usually irreversible and non-stationary. As a consequence, the usual wisdom coming from physics may not be directly applicable to these systems, and the emerging principles (if any) can be quite different.

As the main goal of studying these systems via mathematics is to propose
a set of equations that can be analyzed (analytically or numerically) so as to lead to explanations and predictions, one is trying to reduce the complexity of the initial system by reducing the number of degrees of freedom, e.g., by changing the level of description. However, one must be careful: 
as the analytical (not to speak of rigorous) derivation, as well as the analysis of the higher level equations are complex, one might be tempted to propose ad hoc simplifications that lose some important features of the emerging dynamics. There is therefore a trade-off between what we are able to achieve with the model at hand (i.e. which techniques we have to solve it) and how much it is realistic or predictive (i.e. how to validate the model). These questions are obviously common to any modelling problem, but we would like to focus here on the content of this paper and what we have learned. 
Let us first focus on the main  problems raised from multi-scale approaches and the mathematical tools are available to solve them.

First, the information-theoretic tools described in Section \ref{Eckehard} allow to identify whether the induced dynamics $(Y,\psi)$ at a \emph{given} higher level of description is closed in the sense that the dynamics of the observable(s) associated to $Y$ is a function of these observables only.
This is particularly interesting, as a given application problem is usually related to certain specific aggregate quantities that hence define $Y$.
In physics, for instance, observables usually emerge naturally from the phenomenological knowledge of the system, although in systems such as, e.g., spin glasses, the definition of these observables is not straightforward.
In the context of agent-based models, to recall the example addressed in Section \ref{Sven}, observables of interest are in many (though not all) cases related to aggregations over agent attributes and this defines an associated state space $Y$.
While in Section \ref{Eckehard} the information-theoretic tools were developed in a finite-state, discrete-time setting, it can be extended also to continuous states and time--- see  \cite{Bertschinger2015} for an example. Their application to the mean-field neuronal dynamics might reveal a relation between the history dependence of the derived covariance term and the information flow across levels.

As we have seen, going to a higher level is meant to reduce the number of degrees of freedom,
but this does not necessarily mean that what we have obtained is simpler: the difficulty might be displaced and there might be a price to pay. We have seen, for example, that the change of level can lead a Markovian system to a non-Markovian one. Which methods do we have to handle systems with a virtually infinite memory? A simple case occurs when correlation decay is exponential as in Section \ref{Bruno}; in this case one can propose a Markovian approximation by cutting the memory beyond the time scale of correlation decay.
Another option can be to refine the level of observation in accordance with the relevant model symmetries, as seen in Section \ref{Sven}.
More generally, tools exist in the field of probability theory (variable length Markov chains, chains with complete connections) or statistical physics (Gibbs distributions). Note that the concept of Gibbs distributions allowing left and right conditioning (i.e., on the past and on the future) could be a proper setting to model anticipation mechanisms as well. They can also handle non-stationary problems.

A different and more complicated problem arises if the higher-level state space $Y$ is not given a priori and the task is to find those quantities (observables) which best characterise the dynamical behaviour of the system.
It may be possible to identify quantities for which the dynamics are closed (such as for the specific sub-population structures in Sections \ref{Sven} and \ref{Bruno}), but generally, in more realistic settings, every considerable dimensionality reduction will go with a loss of precision in relation to the original dynamics.
This leads to the question of how such approximations should be evaluated, which generally has to deal with a trade-off between accuracy of the approximate description in relation to its complexity (and the mathematical solution tools available).
Notice that, for agent-based systems, this problem has been addressed in \cite{Lamarche-Perrin2015}.
For instance, when dealing with neuron populations it is not necessarily sufficient to characterize firing rates of all neurons simultaneously, as neurons may have also spatio-temporal correlations which are not explained by rates. 
But then, the question is which correlations? 
Is it sufficient to include pairwise correlations, or do we have to go to higher order to explain the dynamics?
This is again associated with the definition of the space $Y$ in Figure~\ref{fig:main_mathemacs_diagram}.
However, if the higher-level dynamics $\psi$ is not the induced dynamics from the micro-level process, the closure measures described in Section \ref{Eckehard} are not the appropriate tools to evaluate their quality. Thus, the development of suitable methods to deal with this problem will be a topic of future research. Reduction exists, it is closely related to the system under consideration. In physics, observable usually naturally emerge from the phenomenological knowledge of the system, although in systems such as e.g. spin glasses, the definition of these observables is not straightforward. In systems coming from neuroscience, sociology, economics, the situation is even worse especially as it is difficult to find a set of observables that allows a prediction of the system behaviour. For example, in neuroscience, 

\bigskip
\noindent\textbf{Acknowledgement.}
The research leading to these results has received funding from the European Union's Seventh Framework Programme (FP7/2007-2013) under grant agreement no.~318723: \emph{Mathematics of Multi-Level Anticipatory Complex Systems} (MatheMACS). S.B. also acknowledges financial support by the Klaus Tschira Foundation. D.V. acknowledges the support from the Cluster of Excellence Cognitive Interaction Technology 'CITEC' (EXC 277) at Bielefeld University, which is funded by the German Research Foundation (DFG).
 
\bigskip


\begin{thebibliography}{999}
\bibitem{amari:72}
S.~Amari.
"Characteristics of random nets of analog-like elements".
{\em IEEE Trans. Syst. Man and Cybernetics.}, SMC-2(5):643--657,
  1972.

\bibitem{Araki:1999}
H.~Araki.
 {\em {Mathematical Theory of Quantum Fields}}.
Oxford University Press, 1999.

\bibitem{Atay-Roncoroni}
F.~M. Atay and L.~Roncoroni.
"Exact lumpability of linear evolution equations in {B}anach spaces".
{\em MPI-MIS Preprint Series}, 109/2013.
Available at {\em http://www.mis.mpg.de/publications/preprints/2013/prepr2013-109.html}.

\bibitem{Axelrod1997}
R.~Axelrod.
 {The Dissemination of Culture: A Model with Local Convergence and
  Global Polarization}.
 {\em The Journal of Conflict Resolution}, 41(2):203--226, 1997.

\bibitem{Banisch2014dnc}
S.~Banisch.
"The probabilistic structure of discrete agent-based models".
 {\em Discontinuity, Nonlinearity, and Complexity}, 3(3):281--292,
  2014. 
{\em http://arxiv.org/abs/1410.6277}.

\bibitem{Banisch2015springer}
S.~Banisch.
 {\em {Markov Chain Aggregation for Agent-Based Models}}.
 Understanding Complex Systems. Springer, 2015 (in press).

\bibitem{Banisch2015acs}
S.~Banisch and R.~Lima.
"Markov Chain Aggregation for Simple Agent-Based Models on Symmetric
  Networks: The Voter Model".
 {\em Advances in Complex Systems}, 18(03n04):1550011, 2015.
 {\em http://arxiv.org/abs/1209.3902}.

\bibitem{Banisch2012son}
S.~Banisch, R.~Lima, and T.~Ara\'{u}jo.
"Agent based models and opinion dynamics as {M}arkov chains".
{\em Social Networks}, 34:549--561, 2012.

\bibitem{barber-buchinger-etal:06}
M.~Barber, P.~Blanchard, E.~Buchinger, B.~Cessac, and L.~Streit.
"A luhmann-based model of communication, learning and innovation".
 {\em Journal of Artificial Societies and Social Simulation}, 9(4),
  2006.


\bibitem{ben-arous-guionnet:95}
G.~Ben-Arous and A.~Guionnet.
"Large deviations for langevin spin glass dynamics".
{\em Probability Theory and Related Fields}, 102(4):455--509, 1995.

\bibitem{Blanchard:2010}
P.~Blanchard and M.~Hellmich.
"Decoherence in infinite quantum systems".
 {\em Quantum Africa 2010: Theoretical and Experimental Foundations of
  Recent Quantum Technology}, AIP Conf. Proc. 1469:2--15, 2012.

\bibitem{Blanchard:2003}
P.~Blanchard and R.~Olkiewicz.
"Decoherence induced transition from quantum to classical dynamics".
{\em Rev. Math. Phys.}, 15:217--243, 2003.

\bibitem{bouchaud-cugliandolo-etal:96}
J.~P. Bouchaud, L.~F. Cugliandolo, J.~Kurchan, and M.~M{\' e}zard.
"Mode-coupling approximations, glass theory and disordered systems".
 {\em Physica A}, 226:243--273, 1996.

\bibitem{cessac:94}
B.~Cessac.
"Occurence of chaos and {AT} line in random neural networks".
 {\em Europhys. Lett.}, 26(8):577--582, 1994.

\bibitem{cessac:95}
B.~Cessac.
"Increase in complexity in random neural networks".
{\em J. de Physique}, 5:409--432, 1995.

\bibitem{cessac-doyon-etal:94}
B.~Cessac, B.~Doyon, M.~Quoy, and M.~Samuelides.
"Mean-field equations, bifurcation map, and route to chaos in discrete
  time neural networks".
 {\em Physica 74 D}, pages 24--44, 1994.

\bibitem{Cover1991}
T.~Cover and J.~Thomas.
{\em Elements of Information Theory}.
 Wiley-Interscience, New York, 1991.

\bibitem{dauce-etal:98}
E.~Dauc\'e, M.~Quoy, B.~Cessac, B.~Doyon, and M.~Samuelides.
"Self-organization and dynamics reduction in recurrent networks:
  stimulus presentation and learning".
 {\em Neural Networks}, 11:521--33, 1998.

\bibitem{faugeras-maclaurin:13}
O.~Faugeras and J.~M. Laurin.
"Asymptotic Description of Neural Networks with Correlated Synaptic Weights".
 {\em Entropy} {\bf 17} (7), 4701 (2015).


\bibitem{faugeras-touboul-etal:09}
O.~Faugeras, J.~Touboul, and B.~Cessac.
"A constructive mean field analysis of multi population neural
  networks with random synapticweights and stochastic inputs".
 {\em Frontiers in Computational Neuroscience}, 3(1), 2009.

\bibitem{freeman:75}
W.~Freeman.
 {\em Mass Action in the Nervous System}.
Academic Press, New York, 1975.

\bibitem{Geiger1990}
D.~Geiger, T.~Verma, and J.~Pearl.
"Identifying independence in bayesian networks".
 {\em Networks}, 20(5):507--534, 1990.

\bibitem{geman:82}
S.~Geman.
"Almost sure stable oscillations in a large system of randomly coupled
  equations".
{\em SIAM J. Appl. Math.}, 42(4):695--703, 1982.

\bibitem{Gillespie2013}
N.~I. Gillespie and C.~E. Praeger.
"Neighbour transitivity on codes in hamming graphs".
 {\em Designs, codes and cryptography}, 67(3):385--393, 2013.

\bibitem{Horstmeyer-Atay}
L.~Horstmeyer and F.~M. Atay.
"Characterization of exact lumpability of smooth dynamics on
  manifolds".
 {\em MPI-MIS Preprint Series}, 70/2015.
 http://www.mis.mpg.de/publications/preprints/2015/prepr2015-70.html.

\bibitem{jansen-rit:95}
B.~H. Jansen and V.~G. Rit.
"Electroencephalogram and visual evoked potential generation in a
  mathematical model of coupledcortical columns".
 {\em Biological Cybernetics}, 73:357--366, 1995.

\bibitem{Kemeny1976}
J.~G. Kemeny and J.~L. Snell.
 {\em {Finite Markov Chains}}.
 Springer, 1976.


\bibitem{molgedey-schuchardt-etal:92}
L.~Molgedey, J.~Schuchardt, and H.~Schuster.
"Supressing chaos in neural networks by noise".
 {\em Physical Review Letters}, 69(26):3717--3719, 1992.

\bibitem{moynot-samuelides:02}
O.~Moynot and M.~Samuelides.
"{Large deviations and mean-field theory for asymmetric random
  recurrent neural networks}".
 {\em Probability Theory and Related Fields}, 123(1):41--75, 2002.

\bibitem{naude-cessac-etal:13}
J.~{Naud\'e}, B.~{Cessac}, H.~{Berry}, and B.~{Delord}.
"Effects of cellular homeostatic intrinsic plasticity on dynamical and
  computational properties of biological recurrent neural networks".
 {\em Journal of Neuroscience}, 33(38):15032--15043, Oct. 2013.

\bibitem{Pfante2014}
O.~Pfante, E.~Olbrich, N.~Bertschinger, N.~Ay, and J.~Jost.
"Closure measures for coarse-graining of the tent map".
{\em Chaos: An Interdisciplinary Journal of Nonlinear Science},
  24(1):013136, 2014.

\bibitem{Pfante2014a}
O.~Pfante, E.~Olbrich, N.~Bertschinger, N.~Ay, and J.~Jost.
"Comparison between different methods of level identification".
 {\em Advances in Complex Systems}, 17(2):1450007, 2014.

\bibitem{Schelling1971}
T.~Schelling.
"{Dynamic Models of Segregation}".
 {\em Journal of Mathematical Sociology}, 1(2):143--186, 1971.

\bibitem{sompolinsky-crisanti-etal:88}
H.~Sompolinsky, A.~Crisanti, and H.~Sommers.
"Chaos in random neural networks".
 {\em Physical Review Letters}, 61(3):259--262, 1988.

\bibitem{sompolinsky-zippelius:82}
H.~Sompolinsky and A.~Zippelius.
"{Relaxational dynamics of the Edwards-Anderson model and the
  mean-field theory ofspin-glasses}".
 {\em Physical Review B}, 25(11):6860--6875, 1982.


\bibitem{wilson-cowan:72}
H.~Wilson and J.~Cowan.
"Excitatory and inhibitory interactions in localized populations of
  model neurons".
{\em Biophys. J.}, 12:1--24, 1972.


\bibitem{Schlosshauer:2005}
M.~Schlosshauer.
"Decoherence, the measurement problem, and interpretations of quantum mechanics".
 {\em {Rev. Mod. Phys.}, 76. 1267, 2005}. Available at arXiv:quant-ph/0312059.

\end{thebibliography}


\end{document}